\documentclass[conference, review]{IEEEtran}
\usepackage{siunitx}
\usepackage{amsmath,amssymb,amsfonts,mathtools}
\usepackage{dsfont}
\usepackage{ragged2e}
\usepackage[hyphens]{url}
\usepackage{hyperref}
\usepackage{cite}
\usepackage{graphicx}

\graphicspath{{./fig/}}
\DeclareGraphicsExtensions{.pdf,.jpeg,.png}
\usepackage[caption=false,font=footnotesize]{subfig}
\usepackage{array}

\newcommand{\myAuthors}[0]
    {
        Kang Eun Jeon\textsuperscript{1,4}, 
        Sangheum Yeon\textsuperscript{2}, 
        Jinhee Kim\textsuperscript{1,3},
        Hyeonsu Bang\textsuperscript{1}, 
        Johnny Rhe\textsuperscript{1} and 
        Jong Hwan Ko\textsuperscript{1}
    }

\newcommand{\myEmails}[0]
    {
        \{kejeon, 
        shhj9787, 
        a2jinhee,
        bhs1996, 
        djwhsdj, 
        jhko\}@skku.edu
    }

\newcommand{\myAuthorBlock}[0]{
    \author{
        \IEEEauthorblockN{\myAuthors}
        \IEEEauthorblockA{
        \textsuperscript{1}Department of Electrical and Computer Engineering, Sungkyunkwan University, Suwon, Korea \\
        \textsuperscript{2}Department of Semiconductor Engineering, Sungkyunkwan University, Suwon, Korea\\
        \textsuperscript{3}Department of Electrical and Computer Engineering, Duke University, Durham, NC, USA\\
        \textsuperscript{4}Kim Jaechul Graduate School of AI, Korea Advanced Institute of Science and Technology, Seongnam, Korea\\
        \myEmails
            }
        }
    }
\newcommand{\mysubsection}[1]{\vspace{0.5em}\noindent\textbf{#1}}
\newcommand{\mysubsectionNospace}[1]{\noindent\textbf{#1}}
\newcommand{\x}[0]{$\times$}

\newcommand{\figOne}[0]{
    \begin{figure}
        \centering
        \includegraphics[width=.95\linewidth]{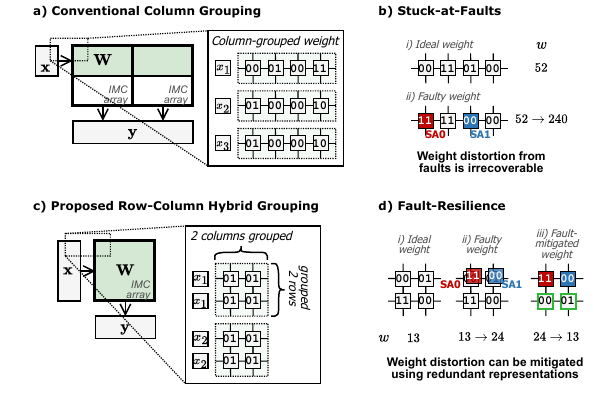}
        \vspace{-1em}
        \caption{Brief comparison of the conventional method versus the proposed one: a) column grouping technique and b) its vulnerabilities against SAFs; c) proposed row-column hybrid grouping and d) its resilience against SAFs. }
        \label{fig1}
        \vspace{-1.5em}
    \end{figure}
}

\newcommand{\figTwo}[0]{
    \begin{figure}
        \centering
        \includegraphics[width=\linewidth]{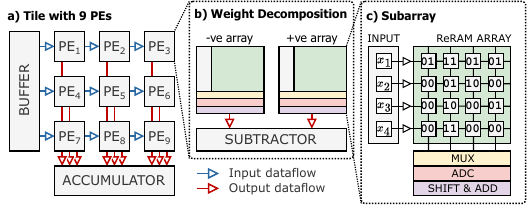}
        \caption{A typical analog IMC architecture and weight mapping method.}
        \label{fig2}
    \end{figure}
}

\newcommand{\figThree}[0]{
    \begin{figure}
        \centering
        \includegraphics[width=0.9\linewidth]{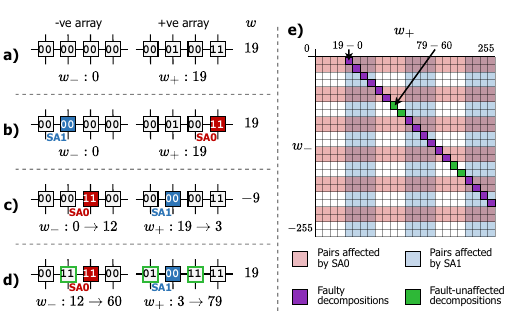}
        \vspace{-1em}
        \caption{Stuck-at-faults and fault-mitigation via weight decomposition. }
        \vspace{-1em}
        \label{fig3}
    \end{figure}
}

\newcommand{\figFour}[0]{
    \begin{figure}
        \centering
        \vspace{1em}
        \includegraphics[width=\linewidth]{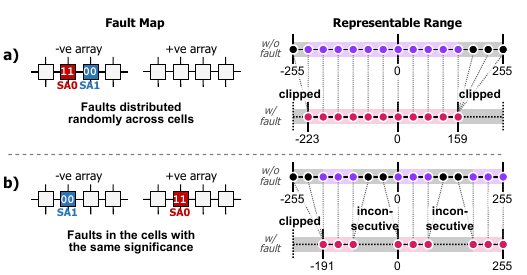}
        \caption{Two sources of fault error: a) clipping; and b) inconsecutivity.}
        \label{fig4}
    \end{figure}
}

\newcommand{\figFive}[0]{
    \begin{figure}
        \centering
        \includegraphics[width=0.9\linewidth]{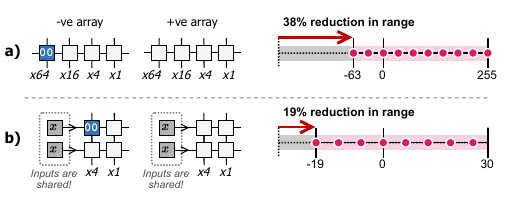}
        \vspace{-1em}
        \caption{Resilience of hybrid grouping against clipping error.}
        \vspace{-1em}
        \label{fig5}
    \end{figure}
}

\newcommand{\figSix}[0]{
    \begin{figure}
        \centering
        \includegraphics[width=0.8\linewidth]{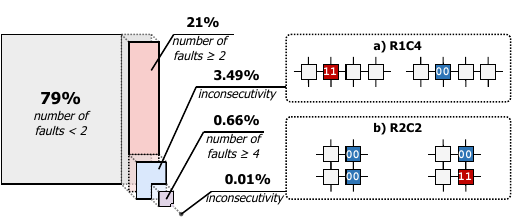}
        \vspace{-1em}
        \caption{Resilience of hybrid grouping against inconsecutivity error.}
        \vspace{-1em}
        \label{fig6}
    \end{figure}
}

\newcommand{\figSeven}[0]{
    \begin{figure*}
        \centering
        \includegraphics[width=\linewidth]{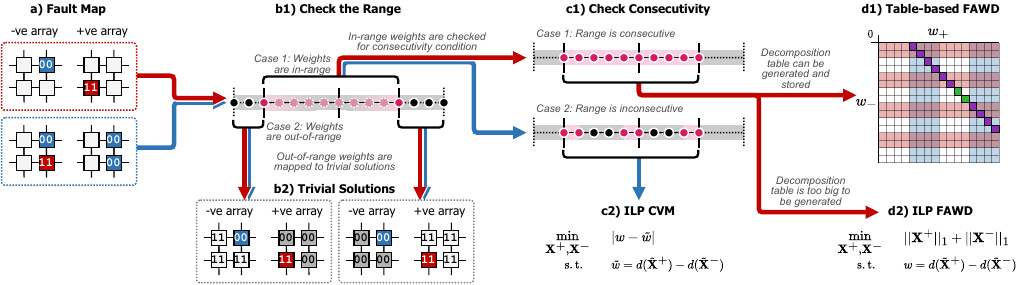}
        \vspace{-1em}
        \caption{Overview of the proposed compilation pipeline.}
        \vspace{-1em}
        \label{fig7}
    \end{figure*}
}

\newcommand{\figEight}[0]{
    \begin{figure}
        \centering
        \includegraphics[width=\linewidth]{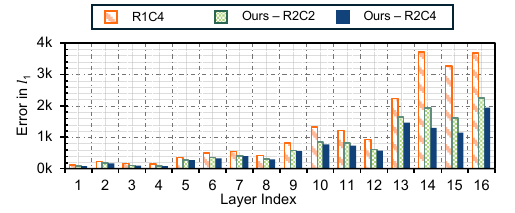}
        \vspace{-2em}
        \caption{Fault + quantization error in ResNet-18.}
        \vspace{-1em}
        \label{fig8}
    \end{figure}
}

\newcommand{\figNine}[0]{
    \begin{figure}
        \centering
        \includegraphics[width=\linewidth]{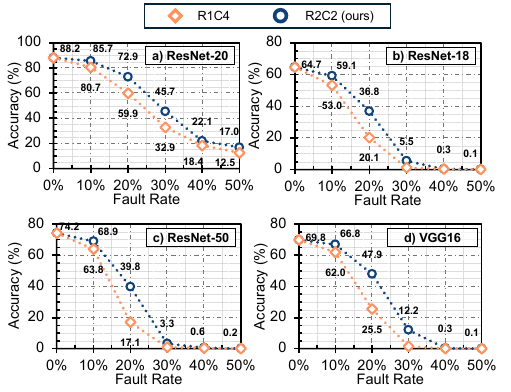}
        \vspace{-2em}
        \caption{Accuracy under varying fault rate.}
        \label{fig9}
    \end{figure}
}

\newcommand{\figTen}[0]{
    \begin{figure}
        \centering
        \includegraphics[width=\linewidth]{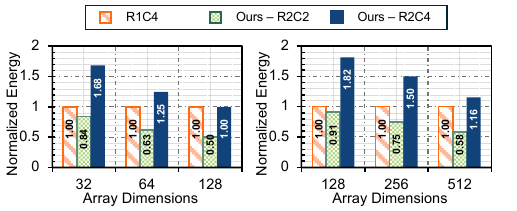}
        \caption{Normalized energy consumption of the proposed against the column grouping method under varying array dimensions.}
        \vspace{-1em}
        \label{fig10}
    \end{figure}
}

\newcommand{\figEl}[0]{
    \begin{figure}
        \centering
        \includegraphics[width=\linewidth]{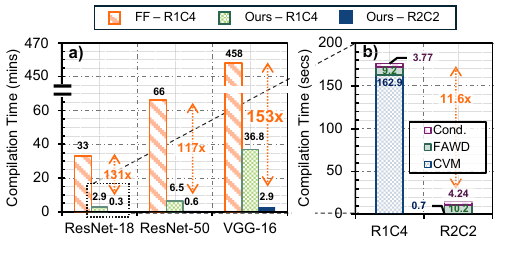}
        \vspace{-2em}
        \caption{Compilation time of the proposed method versus FF.}
        \vspace{-1em}
        \label{fig11}
    \end{figure}
}

\newcommand{\tabOne}[0]{
    \begin{table}[]
        \centering
        \caption{Accuracy evaluation for varying grouping configurations.}
        \includegraphics[width=\linewidth]{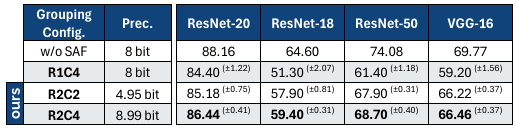}
        \vspace{-2em}
        \label{tab1}
    \end{table}
}

\newcommand{\tabTwo}[0]{
    \begin{table}[]
        \centering
        \caption{Compilation time evaluation for the proposed methods.}
        \includegraphics[width=\linewidth]{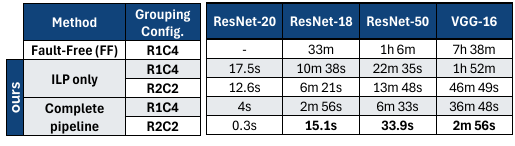}
        \label{tab2}
        \vspace{-2em}
    \end{table}
}

\newcommand{\tabFour}[0]{
    \begin{table}[]
        \centering
        \caption{Perplexity evaluation for varying grouping configurations.}
        \includegraphics[width=\linewidth]{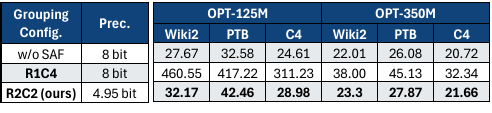}
        \vspace{-3.5em}
        \label{tab4}
    \end{table}
}

\newcommand{\qed}[0]{$\hfill\ensuremath{\Box}$}

\begin{document}
\title{Row-Column Hybrid Grouping for Fault-Resilient Multi-Bit Weight Representation on IMC Arrays}
\myAuthorBlock 

\maketitle
\begin{abstract}
This paper addresses two critical challenges in analog In-Memory Computing (IMC) systems that limit their scalability and deployability: the computational unreliability caused by stuck-at faults (SAFs) and the high compilation overhead of existing fault-mitigation algorithms, namely Fault-Free (FF). To overcome these limitations, we first propose a novel multi-bit weight representation technique, termed row–column hybrid grouping, which generalizes conventional column grouping by introducing redundancy across both rows and columns. This structural redundancy enhances fault tolerance and can be effectively combined with existing fault-mitigation solutions. Second, we design a compiler pipeline that reformulates the fault-aware weight decomposition problem as an Integer Linear Programming (ILP) task, enabling fast and scalable compilation through off-the-shelf solvers. Further acceleration is achieved through theoretical insights that identify fault patterns amenable to trivial solutions, significantly reducing computation. Experimental results on convolutional networks and small language models demonstrate the effectiveness of our approach, achieving up to 8\%p improvement in accuracy, 150\x faster compilation, and 2\x energy efficiency gain compared to existing baselines.

\end{abstract}

\section{Introduction}
The \textit{In-Memory Computing (IMC)} paradigm marks a transformative shift toward non-von Neumann architectures by allowing data processing to occur directly within the memory array \cite{bib:fault-free, bib:ISAAC, bib:PRIME, bib:pipelayer}, thereby minimizing the overhead associated with off-chip data movement \cite{bib:eyeriss}. 
Among various implementations, analog IMC systems based on Resistive Random Access Memory (ReRAM) crossbar arrays have emerged as a particularly promising solution. These systems perform energy-efficient matrix-vector multiplication (MVM)~\cite{bib:PRIME, bib:pipelayer}, a core operation that forms the computational backbone of modern deep learning systems. As such, analog IMC has become a focal point in DNN acceleration and efficient AI research, spearheading cutting-edge investigations in approximate computing, heterogeneous computing, and alternative learning paradigms.

To perform MVM in the analog domain, the weights are stored as conductance values in ReRAM cells; input features are applied as voltages to the word lines, and the resulting bit-line currents naturally multiply-and-accumulate following Ohm’s and Kirchhoff’s laws \cite{bib:raella, bib:low-rank-for-imc}. Here, two key techniques are required to represent high-resolution signed weight values as conductances: \textit{column grouping} and \textit{sign decomposition}. 
Column grouping \cite{bib:column-grouping, bib:partial-sum-quant}  enables multi-bit weight representation using low-resolution ReRAM cells \cite{bib:write-var-1} (Fig.~\ref{fig1}a). Each weight is bit-sliced into low-bit segments, which are then mapped to a group of cells across different columns. 
Sign decomposition \cite{bib:signed-decomp-1, bib:signed-decomp-2}, on the other hand, enables the representation of signed weights in ReRAM-based IMC systems, where conductance values are inherently non-negative. Each weight is decomposed into positive and negative components, which are then mapped to two separate crossbar arrays. The final result is obtained by subtracting the output of the negative array from that of the positive one.

\figOne
Despite the benefits of analog IMC systems, their deployment is impeded by the computational unreliabilities stemming from the device-level analog non-idealities, most notably \textit{Stuck-at-Faults (SAFs)}. 
SAFs are permanent defects that lock a memory cell's resistance state, causing irrecoverable distortions in stored weights \cite{bib:fault_rate} (Fig.~\ref{fig1}b). 
Such distortions induce computational uncertainties that propagate and amplify through the DNN, leading to catastrophic failures \cite{bib:fault-free, bib:saf-impact}.
Fault-aware retraining \cite{bib:fault-tolerant_training, bib:fault-tolerant_training2} adapts DNNs to tolerate faults. However, retraining is often impractical due to unique fault patterns that vary chip-to-chip and limited accessibility to training datasets. Non-retraining methods, such as weight remapping \cite{bib:fault-dcr, bib:vortex}, minimize the impact of faults by relocating sensitive weights to \textit{fault-less} areas but require additional HW peripherals to combat dislocation problems\cite{bib:vortex, bib:PIM-prune}. 

\textit{Fault-Free (FF)} \cite{bib:fault-free} provides a hardware-overhead-free strategy for mitigating SAFs, but incurs significant computational cost during the \textit{offline compilation stage} -- an aspect often overlooked. FF \textit{masks} faults by exploiting redundant representations of weight values, selecting decomposition pairs that preserve the original numerical value despite the presence of stuck-at faults; this problem is known as \textit{fault-aware weight decomposition problem}. However, identifying these representations requires an exhaustive search over a combinatorial space, with complexity that grows rapidly with weight precision. As a result, FF suffers from severe compilation overhead, reporting 33 minutes for  ResNet-18 ($\sim$12M) and up to 8 hours for larger models such as VGG-16 ($\sim$138M). This overhead is particularly problematic in practice, as compilation must be performed per chip, due to the unique SAF patterns that vary chip-to-chip. Furthermore, this is a recurring cost, invoked with every model update or maintenance, hence significantly impeding practical deployment at scale.

This paper addresses the urgent, yet overshadowed, challenges concerning SAFs: (i) computational unreliability and the model accuracy degradation thereof; and (ii) the excessive compilation overhead of FF algorithms,  which hinders the scalable deployment of both the fault-mitigation method and IMC systems. To tackle these issues, we introduce row-column hybrid grouping method (henceforth simply \textit{hybrid grouping}) -- a generalization of the conventional column grouping technique -- paired with a novel compiler pipeline. 
Firstly, the proposed hybrid grouping introduces redundancy in weight representation by grouping memory devices not only across columns but also rows (Fig. \ref{fig1}c). 
This added redundancy can be strategically leveraged to mitigate the effects of SAFs when coupled with FF algorithms (Fig. 
\ref{fig1}d).
Secondly, the proposed compiler pipeline significantly reduces compilation time by reformulating the fault-aware weight decomposition problem as an Integer Linear Programming (ILP) task, enabling the use of efficient and well-established solvers for fast and reliable optimization.
We further accelerate the compilation process by introducing theoretical insights that identify conditions under which fault cases can be simplified, reducing previously complex scenarios to ones that are computationally trivial.
Our experimental evaluations with convolutional networks demonstrate the effectiveness of our approach, achieving up to 8\% points (\%p) improvement in accuracy, 150$\times$ faster compilation time, and 2$\times$ improvement in energy efficiency compared to previous fault mitigation baselines. 

Our key contributions are summarized as follows. 
\begin{itemize}
    \item \textbf{Fault Model \& Fault Errors} - \S\ref{sec:fault-model}:
    Theoretical framework analyzing SAF behavior and fault-induced errors.
    \item \textbf{Row–Column Hybrid Grouping} - \S\ref{sec:hybrid-grouping}: 
    Generalized grouping using extra rows for added SAF resilience.
    \item \textbf{Compiler Pipeline} - \S\ref{sec:compiler}:
    ILP-based compiler for faster search of fault-aware mapping or decomposition pairs.
    \item \textbf{Empirical Evaluations} - \S\ref{sec:exp}:
    Up to 8\%p accuracy boost, and 150$\times$ faster compilation.
    \item \textbf{Open-sourced software}: 
    Full release of framework and compiler for reproducibility (link in \S\ref{sec:setup}).
\end{itemize}

\section{Backgrounds \& Related Works}
\mysubsection{IMC Architecture.} A typical analog IMC architecture consists of ReRAM crossbar arrays that house DNN weights along with peripheral circuitry for MVM operations.
Fig. \ref{fig2}a shows an IMC tile composed of multiple PEs, buffer, and accumulator \cite{bib:SWK_Peng_TCAS2020}. 
To support MVM operation with signed weights, the weights are decomposed into positive and negative parts and stored separately into respective arrays \cite{bib:fault-free, bib:PRIME} as shown in Fig.~\ref{fig2}b. The partial sums from the two arrays are processed through a subtractor to produce the final output. Fig.~\ref{fig2}c details the subarray structure, which includes a ReRAM array, multiplexer, ADC, and shift-and-add circuit. We assume a realistic ReRAM cell of low precision is used and grouped to represent higher precision weights \cite{bib:multi-bit}. To this end, the shift-and-add circuit encodes the significance of each cell and reconstructs the higher precision weight values. 

\figTwo

\mysubsection{Stuck-at-Faults.} ReRAM devices are subject to various analog non-idealities, such as write variation \cite{bib:write-var-1, bib:write-var-2}, process variation \cite{bib:process-var}, and thermal problems \cite{bib:thermal-var}. Among these, SAFs pose a particularly severe threat due to their permanent nature and hence significantly impact DNN inference accuracy \cite{bib:fault-free, bib:saf-impact}. Unlike soft faults, which are transient, SAFs result in memory cells being permanently fixed at a specific resistance state, making them unprogrammable. These defects typically arise during fabrication and can be detected using techniques such as the squeeze-search algorithm \cite{bib:squeeze_search}.

SAFs are categorized as Stuck-At-Zero (SA0) or Stuck-At-One (SA1), corresponding to cells stuck at low or high resistance state, respectively. For example, as shown in Fig. \ref{fig1}b, an SA0 in the Most Significant Bit (MSB) cell and an SA1 in the 2$^\text{nd}$ least Significant Bit (LSB) distort the weight value from 52 to 240, ultimately degrading the model accuracy and potentially rendering the DNN unusable. Reported fault rates include 1.75\% for SA0 and 9.04\% for SA1 in fabricated arrays \cite{bib:squeeze_search}, with other studies indicating up to 11\% of cells may be stuck and untunable \cite{bib:saf-impact}.

\figThree

\mysubsection{Fault-Mitigation Techniques.}
Existing solutions for mitigating SAFs in ReRAM-based IMC systems can be broadly categorized into: (i) retraining-based methods, (ii) hardware-based compensation methods, and (iii) weight remapping methods. 
\textit{Retraining-based methods} adapt model weights to faulty memory patterns via an offline retraining/finetuning stage \cite{bib:fault-tolerant_training, bib:fault-tolerant_training2}. However, they are impractical for edge devices due to chip-specific fault variation and lack of access to proprietary training datasets.
\textit{Hardware-based methods} address SAFs by offloading vulnerable weights to reliable digital hardware or compensating through peripheral circuits. For instance, \cite{bib:fault-free, bib:process-var} routes critical weights to a digital co-processor to preserve accuracy, but this incurs significant energy and area overhead that scales with the offloaded workload.
\textit{Weight remapping strategies} reposition critical weights to avoid defective cells. Methods such as \cite{bib:vortex, bib:fault-tolerant_training2} perform row-wise permutation based on weight importance and fault statistics, while \cite{bib:fault-dcr} applies similar logic to columns. While effective in fault avoidance, remapping disrupts the regular dataflow and introduces alignment issues. As a result, additional peripheral logic (e.g., multiplexers and demultiplexers) is required to correct mismatches between permuted input/output channels.


\mysubsection{Fault-Free.} FF \cite{bib:fault-free} is a promising and tax-free technique that tackles the SAF issues. FF can be viewed as a specialized form of weight remapping that operates entirely within a single weight (i.e., remapping between its grouped memory cells), hence avoiding the need for additional hardware support. FF exploits representation redundancies offered by the positive and negative arrays to identify a combination of $w_+$ and $w_-$ that minimizes the weight distortion for a given faultmap and weight value. 
This principle is illustrated in Fig.~\ref{fig3}. 
First, Fig.~\ref{fig3}a shows a \textit{bitmap} or spatial arrangement of partial weights to collectively represent the target weight. Here, 0 is mapped to the negative array and 19 to the positive array, effectively reconstructing the original weight value of 19.
Fig.~\ref{fig3}b illustrates a case with two faults that are \textit{masked}, meaning they do not alter the stored values, allowing for accurate representation.
Versus, Fig.~\ref{fig3}c presents a different faultmap that causes distortion: here, the negative weight is perturbed from 0 to 12 and the positive weight from 19 to 3.
To correct this, we exploit (i) the redundancy in the weight decomposition and (ii) the principle of masked faults (from Fig.~\ref{fig3}b), to find a combination of $w_+$ and $w_-$ that restores the original weight value as shown Fig.~\ref{fig3}d. The nature of this problem is combinatorial, hence identifying the optimal decomposition pair is challenging, often incurring substantial computational overhead.

Fig.~\ref{fig3}e graphically illustrates the intuition of FF algorithms. Here, a decomposition table is shown, where each cell represents a possible decomposition pair. Pairs affected by SA0 are shaded in red and SA1 in blue. 
Decomposition pairs that represent the original weight value of 19 appear along one diagonal, shaded in either purple or green.
The first stage of FF, known as fault-aware weight decomposition (FAWD), searches for fault-masked decomposition pairs (shaded in green) along this diagonal. If a fault-masked pair cannot be found, FF proceeds to the second stage, closest value matching (CVM), where it searches all remaining pairs in the off-diagonals to minimize weight distortion.

The search for an optimal decomposition pair is computationally intensive, and this issue is only amplified with the large number of parameters in deep learning models. 
Indeed, FF reports compilation time of up to 8 hours for VGG16 -- a model that is now relatively modest in scale. 
Such overhead fundamentally limits the scalability of the FF algorithm, as every model must be compiled per chip based on chip-specific fault maps. This burden will only intensify as modern deep learning models continue to scale in size (e.g., LLMs), and as the demand for higher-precision weights necessitates the use of more memory cells per weight -- resulting in more decomposition choices, a larger search space, and ultimately, higher compilation overhead.

\section{Fault Model \& Fault Errors}
\label{sec:fault-model}

\figFour
To address the above challenges of SAF mitigation solutions, we build upon the works of Shin et al. \cite{bib:fault-free} by proposing a row-column hybrid grouping approach paired with a novel compilation pipeline that jointly improves model performance and reduces compilation time. 

As a foundation, we first develop a theoretical framework to model SAF behavior and uncover new structural insights into how faults distort weight representation, namely in terms of \textit{clipping error} and \textit{inconsecutivity error}. This section presents three key theoretical insights that underpin our contributions, followed by their formal characterization. These findings not only clarify the structure of fault-induced errors but also guide the design of our compiler and optimization strategies.
\begin{itemize}
\item \textbf{Fault Model} (Eqs.~(1), (2)):
We introduce a new linear model for faulty weights that explicitly captures the effect of SAFs. This formulation serves as the basis for reformulating the FAWD and CVM problems as ILPs. \vspace{0.5em}
\item \textbf{Clipping Error} (Theorem 1):  
We prove that if at least one SAF exists within a grouped cells, the resulting error always takes the form of a clipping error -- i.e., a reduction in the representable range of weights -- (see Fig.~\ref{fig4}a). This result allows us to identify weights whose optimal decomposition can be determined trivially when they fall outside the representable range. \vspace{0.5em}
\item \textbf{Inconsecutivity Error} (Theorem 2):  
We show that if all cells corresponding to the same bit significance are affected by SAFs, the resulting error manifests as an inconsecutive error -- i.e., non-contiguous set of representable weights -- (see Fig.~\ref{fig4}b). This insight enables us to select the most appropriate algorithmic strategy for decomposition based on the fault error type. \vspace{0.5em}
\end{itemize}
To formalize the key ideas introduced above, we now provide detailed and rigorous mathematical formulations and proofs.

\mysubsection{Fault Model.} We first define the mathematical construct to detail fault model; more specifically, fault-injection function, faulty bitmap, and faulty weights. Here, bitmap refers to a group of cell where a single DNN weight parameter is loaded. First, we define a function, $f(\cdot)$ that injects faults into our bitmap, $\mathbf{X} \in \mathbb{Z}_{\geq0}^{c \times r}$, to generate a \textit{faulty bitmap}, $\tilde{\mathbf{X}} = f(\mathbf{X},\mathbf{F}_0,\mathbf{F}_1)$:
\begin{equation}
    f(\mathbf{X},\mathbf{F}_0,\mathbf{F}_1) = \underbrace{(1 - \mathbf{F}_0 - \mathbf{F}_1) \odot \mathbf{X}}_{\dot{\mathbf{X}}} + (L-1)\mathbf{F}_0,
    \label{eqn:1}
\end{equation}
where $c$ and $r$ are the number of columns and rows of the bitmap, respectively; $\mathbf{F_0}$ is an indicator matrix SA0 (i.e., $[\mathbf{F}_0]_{i,j} = 1$ iff SA0 exists at $i,j$) and $\mathbf{F_1}$ is an indicator matrix for SA1; $L$ represents the number of levels supported by the memory cell. 
In the first term, $\dot{\mathbf{X}}$, cells affected by the SAFs (both SA0 and SA1) are set to zero, reflecting the influence of SA1. The second term, which accounts for SA1, is then added to capture its influence. Note that $\dot{\mathbf{X}}$ also represents programmable cells (or free variables) that remain unaffected by SAFs. 

Now we can define a \textit{faulty weight}, $\tilde{w} \in \mathbb{Z}$, as follows:
\begin{equation}
    \tilde{w} = d(f(\mathbf{X}^+,\mathbf{F}_0^+,\mathbf{F}_1^+)) - d(f(\mathbf{X}^-,\mathbf{F}_0^-,\mathbf{F}_1^-)),
    \label{eqn:2}
\end{equation}
where $\mathbf{X}^+$ and $\mathbf{X}^-$ are the bitmaps of positive and negative arrays. For notational convenience, we will often use $\tilde{\mathbf{X}}^+$ and $\tilde{\mathbf{X}}^-$ to represent faulty bitmaps (i.e., $\tilde{\mathbf{X}}^+ = f(\mathbf{X}^+,\mathbf{F}_0^+,\mathbf{F}_1^+)$).
$d(\mathbf{X})$ is a decoding function that translates bitmap into an integer weight value defined as $d(\mathbf{X}) = \mathbf{s}\mathbf{X}\mathbf{1}$, where $\mathbf{s}^\top \in \mathbb{Z}^c$ is a significance vector (i.e., $[L^{c-1}, L^{c-2}, \cdots, L^1, 1]$) and $\mathbf{1} \in \mathbb{Z}^r$ is a column vector of ones for summation operation. 
The $\mathbf{1}$ vector ensures that the output of $d(\mathbf{X})$ is a scalar even for $\mathbf{X}$ with $r > 1$. With this setup in place, we now proceed to introduce the two theorems.

\mysubsection{Theorem 1 - Clipping Error.} \textit{If there exists at least one fault in a given faultmap, 
that is $\sum_{\forall i,j} [\mathbf{F}_0]_{i,j} + [\mathbf{F}_1]_{i,j} \geq 1$, 
then the representable range of the faulty bitmap must be strictly less than that of a bitmap without faults:}
\begin{equation}
    \max(d(\tilde{\mathbf{X}})) - \min(d(\tilde{\mathbf{X}})) < \max(d({\mathbf{X}})) - \min(d({\mathbf{X}})), 
    \label{eqn:3}
\end{equation}
\textit{where $\tilde{\mathbf{X}} = \tilde{\mathbf{X}}^+ - \tilde{\mathbf{X}}^-$, the LHS is the representable range of the faulty bitmap, and the RHS that of the ideal bitmap. }
 
\vspace{1em}\noindent\textit{Proof.} We begin by decomposing $d(\tilde{\mathbf{X}})$ into variable and constant components by plugging in (\ref{eqn:1}) to (\ref{eqn:2}):
\begin{equation}
    d(\tilde{\mathbf{X}}) = \underbrace{d(\dot{\mathbf{X}}^+ - \dot{\mathbf{X}}^-)}_{\text{variable component}} + \underbrace{(L-1) \,d(\mathbf{F}_0^+ - \mathbf{F}_0^-)}_{\text{constant component}, C}.
    \label{eqn:4}
\end{equation}
Note that this manipulation is only possible because our decoding function $d(\cdot)$ is linear, hence it is distributive. With this form, we can find the maximum/minimum values of the faulty bitmap by setting the negative/positive variable components to $\mathbf{0}$:
\begin{equation}
    \begin{aligned}
        \max(d(\tilde{\mathbf{X}})) = & \max(d(\dot{\mathbf{X}}^+)) + C,\\
        \min(d(\tilde{\mathbf{X}})) = & -\max(d(\dot{\mathbf{X}}^-)) + C.\\
    \end{aligned}
    \label{eqn:5}
\end{equation}

On the other hand, maximum/minimum values of a bitmap without faults are simply given by:
\begin{equation}
    \max(d(\mathbf{X})) = -\min(d(\mathbf{X})) = d(\mathbf{1}).
    \label{eqn:6}
\end{equation}
Recall that $\dot{\mathbf{X}}$ is a masked version of $\mathbf{X}$ (i.e., element-wise multiplied with a binary matrix where at least one element is 0). Hence it follows that $\max(d(\dot{\mathbf{X}})) < d(\mathbf{1})$. Since a faulty bitmap has at least one fault, this guarantees that either $\max(d(\dot{\mathbf{X}}^+)) < \max(d({\mathbf{X}}))$ or $\max(d(\dot{\mathbf{X}}^-)) < \max(d({\mathbf{X}}))$ is true. Finally, we can reconstruct (\ref{eqn:3}) using (\ref{eqn:5}) and (\ref{eqn:6}), and see that the inequality holds based on the aforementioned relationships. This concludes the proof. \qed

\mysubsection{Theorem 2 - Inconsecutivity Error.} \textit{If faults exist on all cells of the same significance (except for the MSB), and }
\begin{equation}
    \frac{L^i - 1}{L^{i-1} - 1} > 2r
    \label{eqn:7}
\end{equation}
\textit{is true, then its representable range, $\mathbb{S}$, must be inconsecutive: $\exists \tilde{\mathbf{X}} : d(\tilde{\mathbf{X}}) + 1 \notin \mathbb{S}$ where $d(\tilde{\mathbf{X}}) \neq \max(d(\tilde{\mathbf{X}}))$.}

\vspace{1em}\noindent\textit{Proof.} Suppose faults exist on the $i$-th significance cells, where $i \neq c$ and $i \neq 1$. Then, we can rephrase (\ref{eqn:4}) using block matrix notation as follows:
\begin{equation}
    d(\tilde{\mathbf{X}}) = \underbrace{[\mathbf{s}_m, s_i, \mathbf{s}_l]}_{\mathbf{s}}
    \underbrace{[\dot{\mathbf{X}}_m^\top, \dot{\mathbf{x}}_i^\top, \dot{\mathbf{X}}_l^\top]^\top}_{\dot{\mathbf{X}} = \dot{\mathbf{X}}^+ - \dot{\mathbf{X}}^-}
    \mathbf{1} + C
    \label{eqn:8}
\end{equation}
where the first term is $d(\dot{\mathbf{X}})$ 
and $\dot{\mathbf{x}}_i$ the $i$-th column of the faulty bitmap. Since faults exist on all $i$-th significance cells, $\dot{\mathbf{x}}_i$ must be a zero vector. 
Now we can simplify (\ref{eqn:8}) as follows:
\begin{equation}
    d(\tilde{\mathbf{X}}) = 
    \underbrace{\mathbf{s}_m\dot{\mathbf{X}}_m\mathbf{1}}_{\tilde{w}_m} + 
    \underbrace{\mathbf{s}_l\dot{\mathbf{X}}_l\mathbf{1}}_{\tilde{w}_l} + 
    C
\end{equation}
where $\tilde{w}_m$ and $\tilde{w}_l$ are partial weights. Notice that the value of $\tilde{w}_m$ can only change in increments of its smallest significance, $s_{i+1}$ (or $L^{i}$).
Meanwhile, maximum/minimum values of $\tilde{w}_l$ can be found using (\ref{eqn:6}) as $\pm r\sum_{j=1}^{i-1}s_{j}(L-1) $. The expression simplifies to:
\begin{equation}
     \max(\tilde{w}_l) = r(L^{i-1} - 1), \quad \tilde{w}_l = -\max(\tilde{w}_l).
     \label{eqn:10}
\end{equation}

Let us choose an arbitrary $\tilde{\mathbf{X}}$, say $\tilde{\mathbf{X}}_1$, such that its $\tilde{w}_m$ is fixed to some value $a$. Then $\min(d(\tilde{\mathbf{X}}_1))$ is given by $a - \max(\tilde{w}_l)$. Now, suppose we choose another $\tilde{\mathbf{X}}$, $\tilde{\mathbf{X}}_2$, such that its $\tilde{w}_m$ is smaller than $a$ but as close to $a$ as possible, i.e., $\tilde{w}_m = a - s_{i+1}$. 
Then, $\max(d(\tilde{\mathbf{X}}_2))$ will be $a - L^i + \max(\tilde{w}_l)$. 
By definition of consecutivity, the range is consecutive if $\max(d(\tilde{\mathbf{X}}_2)) + 1 \geq \min(d(\tilde{\mathbf{X}}_1))$ is satisfied; otherwise it is inconsecutive. Thus, the condition for inconsecutivity can be expressed as: 
\begin{equation}
    a - L^i + \max(\tilde{w}_l) + 1 < a - \max(\tilde{w}_l).
    \label{eqn:11}
\end{equation}
Substituting (\ref{eqn:10}) into (\ref{eqn:11}) followed by simple algebraic manipulation yields (\ref{eqn:7}), thereby concludes the proof. \qed

\section{Row-Column Hybrid Grouping}
\label{sec:hybrid-grouping}
\mysubsection{Key Idea.}  
The core idea of row-column hybrid grouping is to increase fault resilience by introducing additional representational redundancy when encoding multi-bit weights on ReRAM arrays. While traditional column grouping distributes bit-sliced weight segments across multiple columns, \textit{hybrid grouping} extends this by also grouping rows.
Row grouping is achieved by enforcing shared input voltages across grouped rows. Without loss of generality, consider a group of two rows receiving identical inputs (i.e., $x_1 = x_2$). Let $\mathbf{w}_1, \mathbf{w}_2 \in \mathbb{Z}^c$ denote the weights mapped to these rows, where $c$ is the number of grouped columns. The output of the analog MVM is then:
$\mathbf{w}_1 x_1 + \mathbf{w}_2 x_2 = (\mathbf{w}_1 + \mathbf{w}_2)x_1$.
Although $\mathbf{w}_1$ and $\mathbf{w}_2$ are physically mapped to separate rows, they behave collectively as a single weight due to the shared input. This enables multiple valid decompositions of a target weight into combinations of $\mathbf{w}_1 + \mathbf{w}_2$, thereby creating redundancy in the weight representation.
In the following, we analyze how hybrid grouping enhances fault resilience by revisiting the clipping and inconsecutivity errors previously introduced in our theoretical framework.



\mysubsection{Resilience Against Clipping Error.} Fig.~\ref{fig5} illustrates the impact of faults on the representable range under two grouping configurations. Fig. \ref{fig5}a shows a faultmap for a 1-row, 4-column grouped bitmap (R1C4), while Fig. \ref{fig5}b shows a 2-row, 2-column grouped bitmap (R2C2). In each case, a fault occurs in the MSB cell. In the R1C4 configuration, the representable range is reduced by 38\%, whereas in the R2C2 configuration, it’s only reduced by 18\%. This difference arises because, in the R2C2 configuration, significance is more evenly distributed among the grouped cells. In R1C4, the MSB holds a significance of 64, while in R2C2, there are two MSBs, each with a significance of 4. Consequently, the impact of faults in the proposed hybrid grouping is dampened by the distributed importance across cells, enhancing its fault resilience. However, this fault resilience comes at the cost of precision: with distributed significance, the R2C2 configuration can represent only 31 levels, compared to 256 levels in R1C4.

\mysubsection{Resilience Against Inconsecutivity Error.} Furthermore, the R2C2 configuration is less prone to inconsecutivity error, as it requires four faults rather than two as in R1C4. Fig.~\ref{fig6} shows the probability of inconsecutivity for the two configurations. Note that, with R1C4, inconsecutivity occurs quite frequently with a probability of 3.49\%. Versus, with hybrid grouping, the probability is only 0.01\%, greatly enhancing resilience against inconsecutivity error. Furthermore, resilience against inconsecutivity error will not only improve the model accuracy, but also contribute to reducing compilation time. As finding the decomposition pair for a faultmap with inconsecutivity error requires CMV algorithm, which is computationally more demanding than FAWD algorithm. 


\figFive
\figSix

\figSeven

\section{Compilation Pipeline}
\label{sec:compiler}
\mysubsection{ILP-Fault-Free.} 
To reduce the compilation time and memory overhead of the original FF algorithm, we propose framing it as an ILP problem. This approach allows us to leverage efficient optimization techniques such as the simplex method, and branch-and-bound. First, we define the optimization problem for the FAWD algorithm, which (i) provides a solution only if the fault error can be reduced to zero and (ii) returns the sparsest solution if multiple options exist. With these characteristics, we define the objective as follows:
\begin{equation}
    \begin{aligned}
        \min_{\mathbf{X}^+, \mathbf{X}^-} \quad & ||\mathbf{X}^+||_1 + ||\mathbf{X}^-||_1 \\
        \mathrm{s.t.} \quad & w = d(f(\mathbf{X}^+, \mathbf{F}^+_0, \mathbf{F}^+_1) - f(\mathbf{X}^-, \mathbf{F}^-_0, \mathbf{F}^-_1)) \\
        & 0 \leq x^+_{i,j}, x^-_{i,j} \leq L-1  \quad \forall \, i,j
    \end{aligned}
\end{equation}
where $w$ is the target weight that we aim to represent; and $||\cdot||_1$ is the vector $\ell_1$-norm operator. The objective function minimizes $\ell_1$-norm to find the sparsest bitmap. Since $\mathbf{X}^+$ and $\mathbf{X}^-$ are non-negative, the $\ell_1$-norm operator simply reduces to a simple summation. As previously noted, $d(\cdot)$ and $f(\cdot)$ are linear functions. Consequently, the problem becomes an ILP and can be efficiently solved using ILP solvers. 

On the other hand, the CVM algorithm focuses on minimizing fault error: $\min \, |w-\tilde{w}|$, where $|\cdot|$ is an absolute value function, which is non-linear. We can convert this problem into a linear form by introducing an auxiliary variable $t$:  
\begin{equation}
    \begin{aligned}
        \min_{\mathbf{X}^+, \mathbf{X}^-, t} \quad & t \\
        \mathrm{s.t.} \quad &  -t \leq w - \tilde{w} \leq t \\
        & \tilde{w} = d(f(\mathbf{X}^+, \mathbf{F}^+_0, \mathbf{F}^+_1) - f(\mathbf{X}^-, \mathbf{F}^-_0, \mathbf{F}^-_1)) \\
        & 0 \leq x^+_{i,j}, x^-_{i,j} \leq L-1  \quad \forall \, i,j
    \end{aligned}
\end{equation}
where our optimization variable now includes $t$ and our objective function is minimizing $t$. Thanks to the first constraint, minimizing $t$ is equivalent to minimizing $|w-\tilde{w}|$, just without the absolute value function. Now the problem is linear and can be solved by ILP solvers. 

\mysubsection{Compilation Pipeline.}
Fig. \ref{fig7} presents a graphical illustration of the proposed compilation pipeline, which assembles all aforementioned technical components. Given a faultmap and a target weight, the pipeline aims to identify $\mathbf{X}^+$ and $\mathbf{X}^-$ that minimizes the fault error. Here, we will consider, without loss of generality, two exemplar faultmaps (Fig. \ref{fig7}a): i) a faultmap with consecutive range (red dotted box) and ii) a faultmap with inconsecutivity (blue dotted box). The first stage of our pipeline (Fig. \ref{fig7}b) computes and checks the representable range of a given faultmap, exploiting the insights and equations derived in Theorem 1. If our target weight falls outside the representable range, our solution is trivial. 
The optimal solution that minimized $|w - \tilde{w}|$ would be at either the maximum/minimum value of the range. 
Then, we simply populate one bitmap with 1s and the other with 0s, as we have seen in the proof of Theorem 1. 
If our target weight falls within the representable range, we move on to the second stage.

In the second stage (Fig. \ref{fig7}c), the pipeline checks for the consecutivity of the representable range, based on Theorem 2. If our representable range is inconsecutive, there is a chance the target weight lives in one of the inconsecutive regions;
in such case, the FAWD problem will be unsolvable. Hence, we proceed to solve the problem via the ILP CVM algorithm. On the other hand, if our representable range is consecutive, it is guaranteed that our target weight is representable. Hence, we proceed to solve the problem via the FAWD algorithm. Depending on the hybrid grouping configuration, either Table-based FAWD or ILP FAWD may be employed. If the configuration involves too many cells, decomposition table generation might be computationally intractable. In such a case, ILP FAWD is employed, otherwise, table-based FAWD is preferred.

\section{Experimental Setup}
\label{sec:setup}
\mysubsectionNospace{Implementation Details.}  
We implemented the full compilation pipeline in Python. Gurobi was used to solve ILP problems, while NumPy and Numba were employed to accelerate other components of the pipeline. Additionally, we developed a PyTorch-based simulation platform to assess the impact of SAFs on DNN inference accuracy.  
The complete implementation is available on GitHub\footnote{\href{https://github.com/kejeon/row-col-hybrid-grouping}{https://github.com/kejeon/row-col-hybrid-grouping}}.

\mysubsection{Fault Probabilities.}  
Unless otherwise specified, we assume fault rates of 1.75\% for SA0 and 9.04\% for SA1, as reported in \cite{bib:fault_rate}. The fault distribution is uniform across all bit positions, affecting MSBs and LSBs equally. Our evaluation focuses on memory cells with 1- and 2-bit resolution.

\mysubsection{Models and Datasets.}  
To demonstrate the generality of our framework, we evaluate four CNN architectures: ResNet-20, ResNet-18, ResNet-50, and VGG-16. ResNet-20 is trained and tested on the CIFAR-10 dataset, while the others are evaluated on the ImageNet dataset. We also evaluate our method on compact language models from the OPT family. Model performance is measured via perplexity on text generation tasks using the WikiText-2, PTB, and C4 datasets.

\mysubsection{Quantization.}  
CNN models are trained from scratch and quantized following the AnyPrecision framework \cite{bib:anyprecision}.  

For language models, we apply post-training quantization using GPTQ \cite{bib:gptq}. Symmetric quantization is used, with group size set to full row. Calibration is performed using 128 randomly sampled 2048-token segments from the C4 dataset.

\tabOne
\figEight
\figNine

\section{Experimental Results}
\label{sec:exp}
\mysubsectionNospace{Accuracy Evaluation.} To assess the fault-mitigation capability of the proposed hybrid grouping method, we tested CNN models with various grouping configurations: R1C4, R2C2, and R2C4. The R1C4 configuration represents the traditional column grouping and will serve as our baseline. 

Table~\ref{tab1} summarizes our results. 
The first row shows the ideal accuracy of the DNN models without SAFs. 
With R1C4, significant accuracy drops of 4\%p to 12\%p are observed across all models, indicating high fault vulnerability. 
In contrast, our hybrid configurations mitigate accuracy loss effectively, with R2C4 achieving the best results. Interestingly, R2C2, with 4.95-bit precision, outperforms R1C4 at 8-bit precision, suggesting that the impact of faults outweighs that of quantization. 

Supporting this claim, Fig.~\ref{fig8} illustrates the layer-wise error of ResNet-18, measured with the $\ell_1$ loss function. It can be observed that with the proposed hybrid grouping configuration, the combined error from both quantization and faults is significantly reduced—by up to approximately 50\%—thereby improving overall accuracy. Besides, Fig.~\ref{fig9} presents the accuracy performance under varying fault occurrence rates; here, the occurrence ratio of SA0 and SA1 is fixed to 1.75:9.04, and the probability of SAF occurrence rate is controlled. Again, we can observe the proposed approach consistently outperforms the traditional column grouping counterpart. 

\tabTwo
\figEl

\mysubsection{Compilation Time Evaluation.} We evaluated the compilation time of the proposed pipeline on an Intel Xeon Silver 4210 processor, matching the setup in the original FF paper \cite{bib:fault-free}. All evaluations were conducted with a single processor thread unless specified otherwise. Table~\ref{tab2} summarizes results across different DNN models and grouping configurations. The first row of the table shows the compilation times reported in FF with column grouping. The second and third rows show measurements for FF enhanced with our ILP formulation (i.e., without range or consecutivity checking stages), while the fourth and fifth rows present results for the full pipeline.

Fig.~\ref{fig11} focuses on the comparisons between FF, our baseline, and the complete pipeline. As shown in Fig.~\ref{fig11}a, the results indicate a significant speed-up of up to 153x on the VGG-16 network with the R2C2 configuration. This improvement extends to other models, with the proposed method achieving at least a 100x speed-up across the board. Additionally, the pipeline is compatible with traditional column grouping, delivering a notable 10x speed-up over our baseline—a result that could be further enhanced with multithreading.

The synergy between the hybrid grouping and the compilation pipeline is studied in Fig.~\ref{fig11}b, where the compilation times are broken down into three operations: i) range and consecutivity condition checking (Cond.), ii) FAWD, and iii) CVM.  In the R1C4 configuration, the CVM stage dominates the compilation time, accounting for 90\% of the total. In contrast, for the R2C2 configuration, CVM time is reduced to just 0.7 seconds. This result aligns with Theorem 2: inconsecutivity errors are rare in R2C2, allowing the majority of weights to be efficiently handled by FAWD.

We also evaluated the compilation time for the R2C4 configuration, which requires ILP-FAWD. For this experiment, we used 4 threads, and the results show that R2C4 takes 15 mins to compile ResNet-18 and 34 mins for ResNet-50. Although slower than other configurations, this is still faster than the original FF approach. 
Moreover, FF alone fails to compile R2C4, as the corresponding decomposition table becomes prohibitively large, whereas our method enables this configuration regardless of the compilation time constraints.

\tabFour
\mysubsection{Language Model Evaluation. }
We evaluate the fault resilience of the proposed hybrid grouping method on two language models: OPT-125M and OPT-350M. Larger models are not considered, as their memory footprint exceeds the capacity of typical IMC arrays.
Table~\ref{tab4} summarizes the perplexity results on WikiText-2, PTB, and C4 datasets, averaged over 10 trials. Across all models and datasets, the R2C2 configuration consistently achieves lower perplexity than R1C4, indicating better fault tolerance.  

Notably, R1C4 exhibits significant deviation from the baseline (i.e., SAF-free) perplexity, especially for smaller models. For instance, on WikiText-2, the R1C4 configuration yields a perplexity of $\sim$460 for OPT-125M -- compared to a baseline of 27.67 -- indicating severe model performance degradation. In contrast, R2C2 achieves a perplexity of 32.17, remaining much closer to the original performance.  
These results highlight that SAFs can cause catastrophic failures, especially in smaller language models, while the proposed hybrid grouping effectively mitigates such failures.

The evaluation of the large language models is made possible only through our accelerated compilation pipeline. Leveraging multi-threading, the compilation time for the OPT-350M model takes under two to three minutes. This demonstrates the scalability of our framework and showcases its potential to bring FF-style fault mitigation to larger models and more practical IMC deployments. 

\mysubsection{Hardware Evaluation. }
We evaluate the hardware performance of CNN models using the proposed hybrid grouping. To this end, we built a simulator around NeuroSIM~\cite{bib:neurosim}  and ConvMapSIM~\cite{bib:convmapsim} to measure the energy consumption of the proposed method. For convolutional weight mapping, we selected kernel splitting, the default mapper in NeuroSIM. While known for its energy efficiency, it often suffers from low row utilization with larger IMC arrays, leading to performance drops. We assessed energy consumption for both ResNet-20 and ResNet-18 networks across various array sizes. Fig.~\ref{fig10} plots the normalized energy consumption of the R2C2 and R2C4 configurations compared to the baseline R1C4 method. 

The R2C2 configuration demonstrates significant energy savings for both networks, reducing energy consumption by up to 50\%. This is noteworthy given that R2C2 also showed substantial accuracy gains over R1C4. We attribute this boost in energy efficiency to the improved array utilization, especially in the shallower layers of the CNN models. These layers typically have a small number of input channels, leading to underutilization of rows in conventional column grouping schemes. In contrast, the proposed hybrid grouping reduces column usage while increasing row utilization, thereby enhancing overall array utilization and energy efficiency. 


\figTen

\section{Conclusion}
In this work, we proposed row–column hybrid grouping -- a novel multi-bit weight representation technique -- paired with an ILP-based compilation pipeline for fault mitigation in analog IMC systems. Our method achieves up to 150× faster compilation, 2× better energy efficiency, and 8\%p higher accuracy on CNNs. We also demonstrate its effectiveness on language models, maintaining superior fault-tolerance compared to the column-grouping baseline while keeping compilation time under three minutes. Beyond these practical gains, we establish a new theoretical framework that formalizes SAF-induced errors through clipping and inconsecutivity. Moreover, to support reproducibility, we open-sourced our entire implementation, including the compiler and fault simulation pipeline. Together, these contributions advance scalable, fault-resilient analog IMC across both vision and language domains.

\section*{Acknowledgment}
This work was partly supported by the National Research Foundation of Korea (NRF) grant 
(No. RS-2024-00345732);
the Institute for Information \& communications Technology Planning \& Evaluation (IITP) grants 
(RS-2020-II201821,
RS-2019-II190421, 
RS-2021-II212052,
RS-2021-II212068,
RS-2025-02217613,
RS-2025-10692981,
RS-2025-25442569); 
the Technology Innovation Program 
(RS-2023-00235718, 
23040-15FC) funded by the Ministry of Trade, Industry \& Energy (MOTIE, Korea) grant (1415187505);
Samsung Electronics Co., Ltd (IO230404-05747-01).

\bibliographystyle{IEEEtran}
\bibliography{IEEEabrv,mybibfile}

\end{document}